\documentclass[conference, 10pt]{IEEEtran}

\usepackage{cite}
\usepackage{url}
\usepackage{graphicx}
\usepackage{color}
\usepackage{placeins}
\usepackage{float}
\usepackage{tabularx,colortbl}
\usepackage{ifthen}
\usepackage{multirow}
\usepackage{amsmath} 
\usepackage{booktabs}
\usepackage{caption} 

\makeatletter

\newcounter{author}
\renewcommand{\author}[2][]{
   \stepcounter{author}
   \@namedef{author@\theauthor}{#2}
   \@namedef{authorlabel@\theauthor}{#1}
}

\newcounter{address}
\newcommand{\address}[2][]{
   \stepcounter{address}
   \@namedef{address@\theaddress}{#2}
   \@namedef{addresslabel@\theaddress}{#1}
}

\newcommand{\alsep}{and}

\def\newmaketitle{\par%
  \begingroup%
  \normalfont%
  \def\thefootnote{}
  \def\footnotemark{}
  \let\@makefnmark\relax
  \footnotesize
  \footnotesep 0.7\baselineskip
  \normalsize%
  \twocolumn[\thenewmaketitle\@IEEEaftertitletext]%
  \if@IEEEusingpubid
     \enlargethispage{-\@IEEEpubidpullup}%
  \fi
  \endgroup
  \setcounter{footnote}{0}\let\maketitle\relax\let\@maketitle\relax
  \gdef\@thanks{}%
  \let\thanks\relax}

\def\thenewmaketitle{
  \newpage
  \begin{center}%
    \vskip0.2em{\Huge\@IEEEcompsoconly{\sffamily}\@IEEEcompsocconfonly{\normalfont\normalsize\vskip 2\@IEEEnormalsizeunitybaselineskip
   \bfseries\large}\@title\par}\vskip1.0em\par%
    \vspace{1ex}
    \newcounter{c@author}
    \newcounter{c@tmp}
    \ifthenelse{\value{author}=2}{%
      \newcommand{\liand}{ and }}{%
      \newcommand{\liand}{, and }}
    \ifthenelse{\value{address}<2}{%
      \@nameuse{author@1}%
      \stepcounter{c@author}%
      \whiledo{\value{c@author}<\value{author}}{%
        \setcounter{c@tmp}{\value{author}}%
        \addtocounter{c@tmp}{-\value{c@author}}%
        \ifthenelse{\value{c@tmp}=1}{%
          \renewcommand{\alsep}{\liand}}{\renewcommand{\alsep}{, }}%
        \stepcounter{c@author}\alsep \@nameuse{author@\thec@author}}\\%
    }
    {
      \@nameuse{author@1}${}^{(\ref{\@nameuse{authorlabel@1}})}$%
      \stepcounter{c@author}%
      \whiledo{\value{c@author}<\value{author}}{%
      \setcounter{c@tmp}{\value{author}}%
      \addtocounter{c@tmp}{-\value{c@author}}%
      \ifthenelse{\value{c@tmp}=1}{%
        \renewcommand{\alsep}{\liand}}{\renewcommand{\alsep}{, }}%
      \stepcounter{c@author}\alsep \@nameuse{author@\thec@author}%
        ${}^{(\ref{\@nameuse{authorlabel@\thec@author}})}$%
      }
    }
    \vspace{0.2ex}

    \ifthenelse{\value{address}>0}{%
      \ifthenelse{\value{address}=1}{
        {\@nameuse{address@1}}
      }
      {
        \newcounter{c@address}

        \begin{center}
        \whiledo{\value{c@address}<\value{address}}
        {
          \refstepcounter{c@address}
            ${}^{(\thec@address)}$\,%
              \label{\@nameuse{addresslabel@\thec@address}}%
              \@nameuse{address@\thec@address}\\ %
        }
        \end{center}
      } 
    }
    {
      \relax
    }
  \end{center}
}

\makeatother

\title{Estimating Rural Path Loss with ITU-R P.1812-7: Impact of Geospatial Inputs}

\author[org1]{Mathieu Châteauvert}
\author[org1]{Jonathan Ethier}
\author[org1]{Adrian Florea}

\address[org1]{Communications Research Centre Canada (CRC), Ottawa, Ontario, Canada \\ \{mathieu.chateauvert, jonathan.ethier, adrian.florea\}@ised-isde.gc.ca}

\begin{document}

\newmaketitle

\begin{abstract}
Accurate radio wave propagation modeling is essential for effective spectrum management by regulators and network deployment by operators. This paper investigates the ITU-R P.1812-7 (P.1812) propagation model's reliance on geospatial inputs, particularly clutter information, to improve path loss estimation, with an emphasis on rural geographic regions. The research evaluates the impact of geospatial elevation and land cover datasets, including Global Forest Canopy Height (GFCH), European Space Agency WorldCover, and Natural Resources Canada LandCover, on P.1812 propagation model prediction accuracy. Results highlight the trade-offs between dataset resolution, geospatial data availability, and representative clutter height assignments. Simulations reveal that high-resolution data do not always yield better results and that global datasets such as the GFCH provide a robust alternative when high-resolution data are unavailable or out-of-date. This study provides a set of guidelines for geospatial dataset integration to enhance P.1812's rural path loss predictions.
\end{abstract}


\section{Introduction}
Developing reliable methods for radio wave propagation prediction is essential for addressing the challenges of modern communication network design, especially in rural areas where environmental factors can significantly impact signal propagation \cite{SAFE_PAPER}. Effectively predicting how radio waves propagate across diverse terrains enables improved network planning, leading to more reliable communication services.

To estimate path loss between transmitting and receiving antennas, International Telecommunication Union - Radiocommunication (ITU-R) P.1812-7 (herein referred to as P.1812) \cite{P1812} leverages above-ground clutter information to enhance the accuracy of its predictions across diverse geographical environments. Despite its sophistication, its implementation can be complex, primarily due to its reliance on clutter category information that is not explicitly defined in the current ITU-R Recommendation \cite{P1812}. Providing specific guidelines for categorizing zones based on their clutter characteristics could further improve its operational effectiveness.

This paper evaluates approaches to improve the prediction accuracy of P.1812 by assessing different geospatial datasets (geodata) and clutter classification methodologies. It explores methods for assigning clutter heights, the impact of geodata resolution, and the evaluation of various land cover datasets.

\section{Background}

\subsection{ITU-R P.1812-7}

The P.1812 model, designed for predicting coverage in point-to-area services, directly accounts for the clutter information above-ground, specifically buildings and foliage, to improve the accuracy of path loss predictions.  It calculates the path loss using the path profile between the transmitter and receiver antennas. A path profile consists of three sets of multiple evenly spaced points with each point containing information on: (1) the distance from the transmitter; (2) the terrain height above sea level; and (3) the representative clutter height. Representative clutter heights refer to the actual or typical heights assigned to the clutter to model how they interact with radio waves. The first two sets of necessary inputs are easily calculated — the distance can be obtained using the GeoPy library \cite{geopy} and the terrain elevation derived from a Digital Terrain Model (DTM) database — it is however non-trivial to determine the third input parameter, the clutter category at every point. Without incorporating clutter into the path profile, diffraction loss in cluttered environments is likely to be underestimated compared to a combined representation of terrain and clutter, emphasizing the importance of modeling above-ground obstructions.

Recommendation P.1812 suggests two methods for incorporating surface data to enhance the estimation of path loss, thereby addressing the third input requirement for the model. Section 3.2.1 of \cite{P1812} constructs a surface profile using representative clutter heights based on clutter categories derived from land cover datasets. These datasets categorize the physical features of Earth's surface, such as vegetation, urban infrastructure, and water bodies. However, land cover datasets may not provide the detailed classification required for the P.1812 model, as they often do not break down urban infrastructure into multiple subcategories. In some cases, land cover datasets may differentiate certain categories, such as trees, into more specific types such as deciduous or coniferous.

The ITU-R Recommendation provides default representative clutter heights, detailed in Table \ref{tab:clutter_categories}. Users can employ these values in cases where specific height data is unavailable. However, geodata corresponding to the clutter categories listed in Table \ref{tab:clutter_categories} are not always readily accessible.

\begin{table}[h!]
\caption{P.1812 Clutter Categories and Representative Clutter Height.}
\label{tab:clutter_categories}
\centering
\begin{tabular}{lc}
\toprule
\textbf{Clutter Category} & \textbf{Default Representative Clutter Height (m)} \\ \midrule
Water / Open / Rural & 0 \\ 
Suburban & 10 \\ 
Urban / Trees / Forest & 15 \\
Dense Urban & 20 \\ \bottomrule
\end{tabular}
\end{table}


Section 3.2.2 of \cite{P1812} states that surface height profiles can be used as inputs to the model. In this study, height profiles were extracted from Digital Surface Model (DSM) datasets, which represent the actual height of above-ground obstacles.

\section{Geodata Methodology}
\label{sec:methodology}
This section details the various methods employed to create path profiles from the geodata.

\subsection{Using Elevation Datasets}
High-resolution elevation datasets that provide detailed height information have been available in recent years, including in Canada (CAN) \cite{HRDEM} and the United Kingdom (UK) \cite{UK_DTM_DSM}. Those datasets include DSM and DTM maps with a resolution of 1 m. By calculating the difference between DSM and DTM, the height of clutter above the terrain can be determined, which, henceforth, will be referred to simply as the Height Above Ground (HAG). 
Several methods can be used to incorporate the HAG as an input to the P.1812 model. The HAG can be directly used ``as is``; however, as mentioned in \cite{P1812}, ``spacing less than 10 m could lead to an overestimation of [path] loss because the modeling might capture individual obstacles instead of broader surface features``. To address the issue, this paper explores resampling HAG data using various techniques, such as bi-linear interpolation, averaging, or selecting the maximum value, to mitigate the risk of overestimation and align with the ITU-R Recommendation.

Since trees are the primary propagation obstacles in rural areas, another elevation dataset tested was the Global Forest Canopy Height (GFCH) \cite{GFCH}, which has a spatial resolution of 30 m and provides tree height data worldwide. It can be used directly as an input to the model without additional processing.

\subsection{Using Land Cover Datasets}
The European Space Agency WorldCover 2021 (ESA) \cite{esaworldcover} is a land use raster dataset of the entire globe classified at a 10 m resolution. Key land categories, such as \textit{built-up} and \textit{tree}, are provided to inform the clutter characteristics. To determine clutter heights for each pixel, two methods were employed: (1) using the default representative clutter heights suggested by Recommendation P.1812 as a baseline (see Table \ref{tab:clutter_categories}), and (2) calculating statistical measures such as the mean, median (50th percentile), and 75th percentile from the HAG over the given region. These extracted statistics were then used for path profile creation (inputs to the P.1812 model) to evaluate their impact on accuracy.

Other land use datasets, such as Natural Resources Canada LandCover 2020 (NRCan) \cite{CanadaLandCover2020} and Open Street Map Land Usage (OSM) \cite{OpenStreetMap} were also experimented with to determine where discrepancies in land cover classifications or resolutions might improve path loss predictions. The same algorithm described above was applied to assign the heights, involving either the use of default values or explicitly using the heights from the HAG. NRCan includes multiple tree types, allowing for assigning different tree heights to the different tree categories, compared to ESA which has a single tree category. Meanwhile, the \textit{Key:landuse} attribute in OpenStreetMap (OSM), which primarily describes land use by humans, was utilized to extract the wooded areas (\textit{landuse=wood}) and assign appropriate heights to this feature.

\section{Model Evaluation and Analysis}
To determine the most effective approach for path loss estimation, maps were generated from both elevation data and land cover datasets outlined in Section \ref{sec:methodology}. These maps were then utilized by P.1812 to produce predictions, which were subsequently compared against path loss measurements.

\subsection{Path Loss Measurements}
Seven rural path loss measurement datasets were used to validate the different clutter classification approaches. Measurements collected by CRC (Ottawa), NetScout (Laurentides, Peterborough, New Brunswick) \cite{NetScout} as well as Ofcom (Boston, Scar Hill, Merthyr Tydfill) \cite{OFCOM_DATA} were considered. An overview of the data is provided in Table \ref{tab:measurements}. For this study, the datasets were strategically split into three distinct groups: High-Rx CAN, Low-Rx CAN, and Low-Rx UK. Each group represents a unique measurement setup, with differences in geospatial characteristics, leading to variations in data collection conditions. The High-receiver (Rx) group, in particular, warrants special consideration due to both antennas being positioned above the clutter. The year of measurement aligns closely with the collection period of the elevation data, ensuring no significant temporal difference between the two.

The regions all feature a mix of flat and gently rolling terrain. The landscape is characterized by a combination of forests, agricultural fields, and water bodies. The transmitter (Tx) antennas vary significantly in height, spanning from 17 to 107 m. Meanwhile, the receiver antennas are positioned below 3.5 m, except for the High-Rx CAN group, which ranges from 10 to 16 m. The nature of the measurement campaign, which involved stopping and raising the masts at multiple locations, explains the lower number of measurements (131).

\begin{table}[t] 
    \centering
    \caption{Path Loss Measurement Datasets.}
    \label{tab:measurements}
    \begin{tabular}{llcc}
        \toprule
        \textbf{Dataset} & \textbf{Region} & \textbf{Frequency} & \textbf{Number of Records} \\
        & & \textbf{(MHz)} & \textbf{Thousands}\\ 
        \midrule
        High-Rx CAN      & Ottawa          & 3875   & 0.131 \\ \midrule
                         & Laurentides     & 755    & 6.9 \\
        Low-Rx CAN           & New Brunswick   & 3599   & 43.8 \\
                         & Peterborough    & 1945   & 18.3 \\ \midrule
                       & Boston          & \multirow{3}{*}{\begin{tabular}[c]{@{}c@{}}449, 915, \\ 1802, 2695, \\ 3602, 5805\end{tabular}} & 1100 \\
        Low-Rx UK                & Scar Hill       &        & 740 \\
                         & Merthyr Tydfil  &        & 1103 \\
        \bottomrule
    \end{tabular}
\end{table}

Path loss simulations were performed, and the Root Mean Square Error (RMSE) was calculated to compare the simulated values with the measured ones. To account for discrepancies in the number of measurements across datasets, the average RMSE was calculated for each group. The dataset yielding the lowest average RMSE was considered the most accurate. 
\begin{table*}[!t] 
    \centering
    \caption{Comparison of RMSEs across different resolutions and methods using elevation data.}
    \label{tab:elevation_rmse}
    \begin{tabular}{lcccccccccc}
        \toprule
        & \multicolumn{3}{c}{GCFH} & \multicolumn{4}{c}{HAG} & \multicolumn{3}{c}{HAG (Bi-linear resampling)} \\
        \cmidrule(lr){2-4} \cmidrule(lr){5-8} \cmidrule(lr){9-11}
        Sampling Resolution (m) & 10 & 30 & 100 & 1 & 10 & 30 & 100 & 10 & 30 & 100 \\
        \midrule
        High-Rx CAN & 13.9 & 10.6 & \textbf{7.8} & 10.1 & 9.4 & 9.6 & 11.7 & 11.1 & 13.4 & 17.8 \\
        Low-Rx CAN & 10.4 & 10.3 & 11.3 & 15.3 & 12.4 & 8.8 & \textbf{7.4} & 10.4 & 8.2 & 9.6 \\
        Low-Rx UK & 12.2 & \textbf{11.3} & \textbf{11.3} & 13.9 & 14.8 & 14.9 & 15.3 & 14.6 & 14.2 & 14.7 \\
        \bottomrule
    \end{tabular}
\end{table*}

\begin{table*}[!t]
    \centering
    \caption{Comparison of RMSEs across different resolutions and methods using ESA WorldCover.}
    \label{tab:esa_rmse}
    \begin{tabular}{lccccccccc}
        \toprule
        & \multicolumn{3}{c}{Tree Heights=15m (Default)} & \multicolumn{3}{c}{Tree Heights=Mean} & \multicolumn{3}{c}{Tree Heights=75th pct.} \\
        \cmidrule(lr){2-4} \cmidrule(lr){5-7} \cmidrule(lr){8-10}
        Sampling Resolution (m) & 10 & 30 & 100 & 10 & 30 & 100 & 10 & 30 & 100 \\
        \midrule
        High-Rx CAN & 12.5 & 12.8 & 13.5 & 16.1 & 16.5 & 17.5 & 13.2 & 12.3 & \textbf{11.4} \\
        Low-Rx CAN & 16 & 12.1 & \textbf{7.5} & 14.1 & 10.3 & 7.6 & 16.9 & 13.1 & 8.2 \\
        Low-Rx UK & 13.4 & 11.0 & \textbf{9.0} & 11.1 & 9.4 & 10.2 & 13.3 & 10.9 & 9.1 \\
        \bottomrule
    \end{tabular}
\end{table*}

For simulations, some parameters were held constant to ensure the consistency of comparative analyses. The terrain elevation was sourced from the Medium Resolution Digital Elevation Model (MRDEM) \cite{MRDEM} in Canada, which has a resolution of 30 m and the high-resolution elevation data \cite{UK_DTM_DSM} in the UK, which was resampled to 30 m. Within the context of this work, the P.1812 time percentage \textit{p} and location percentage \textit{pL} are both set to 50\%. During the experimentation phase, multiple sampling resolutions were tested — 10 m, 30 m, 50 m, and 100 m — to identify the optimal path profile resolution, despite the ITU-R Recommendation suggesting 30 m. When a resampling was performed, the sampling resolution matched the map resolution.

\subsection{Elevation Data Results and Analysis}

At first glance, as shown in Table \ref{tab:elevation_rmse}, it is apparent that the resolution of the dataset has a substantial effect on the RMSEs outcomes. Variations in map resolution, and thus, data granularity, directly influence the accuracy of calculations as the data may inadvertently capture individual obstacles rather than a more generalized surface profile. While a finer resolution may technically provide additional information, it does not guarantee improved performance of the P.1812 model. 

Despite having a resolution of 30 m, the GFCH dataset can outperform the high-resolution HAG model under certain conditions, as seen in Table \ref{tab:elevation_rmse}. This confirms that the performance of P.1812 is not dependent on higher resolution. Although it may seem beneficial to use ``real heights`` and the highest possible resolution, this approach might not yield optimal results. This is primarily due to the inherent design of P.1812, which can exhibit sensitivity to over-predicting path loss when utilizing high-resolution data. Additionally, the availability of high-resolution datasets is often limited, especially in the Canadian context \cite{HRDEM}. In such instances, the GFCH dataset emerges as a viable alternative, since it provides global coverage. Moreover, the applicability of datasets can vary significantly based on the intended application. For instance, geodata that is adequate for an elevated base station or applicable in one country might not yield optimal results for lower Rx or when used in a different country. The nuanced performance across the different path loss measurement simulations emphasizes the need for context-specific data selection for geographies, frequency and application type.

The examination of Table \ref{tab:elevation_rmse} also reveals that utilizing HAG often yields superior results compared to resampling methods. This could be attributed to the precision gained from describing the path profile between the Tx and Rx locations, which allows for the accurate identification of major obstacles. When resampling techniques such as bilinear, averaging or maximum value selection are applied, obstacle information can be over-attenuated. Moreover, while Table III only presents results for bilinear resampling, averaging methods yield similar results. In contrast, maximum resampling consistently underperforms relative to the other two techniques.

\subsection{ESA WorldCover Results and Analysis}
In the analysis of clutter heights using the land cover dataset, in the case of ESA, several observations emerge. Relying on default tree heights (i.e. 15 m), provides a practical alternative, as shown in Table \ref{tab:esa_rmse}. The results also show that using representative clutter heights derived from elevation datasets might not necessarily outperform the default values. Also, the chosen algorithm for extracting tree heights (using the median or 75th percentile) can significantly influence the results. Further, the RMSEs using the 75th percentile of the tree heights often align with those from default values, as the 75th percentile of the tree height frequently approximates 15 m. For instance, in the UK, Merthyr Tydfil (16 m), Boston (14 m), and Scar Hill (13 m) illustrate this similarity.

\begin{figure}[ht]
    \centering
    \includegraphics[width=0.4\textwidth]{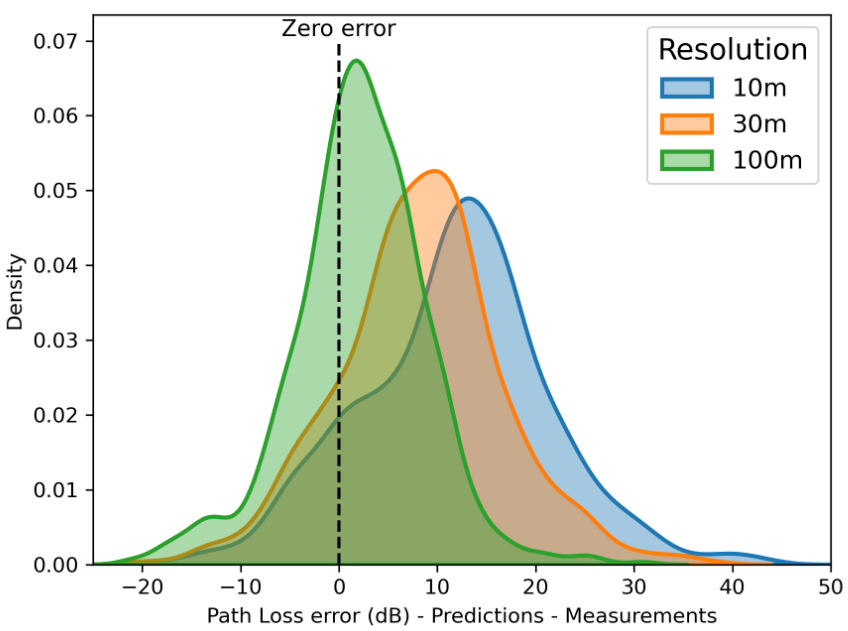}
    \caption{Distribution of the prediction error on the Canadian dataset using the default tree height for different resolution.}
    \label{fig:hist_plot}
\end{figure}

As demonstrated in Table \ref{tab:esa_rmse} and in Fig. \ref{fig:hist_plot}, employing higher resolution does not correlate with a lower RMSE. Specifically, when examining the Canadian dataset and using the default tree heights, the median error is 13.7 dB at a resolution of 10 m, compared to only 2.7 dB at a resolution of 100 m. A median error closer to 0 dB is advantageous, as it indicates predictions align closely with the measured data and introduce no significant bias. A high positive median error suggests that the model tends to over-predict path loss. To address this over-prediction, employing a lower resolution can yield more accurate outcomes.

\subsection{Other Land Use Datasets Results and Analysis}

\begin{table}[t]
    \centering
    
    \setlength{\tabcolsep}{4pt}
    \caption{Comparison of RMSEs for various Land Use Datasets (Resolution = 100m).}
    \label{tab:rmse_other}
    \begin{tabular}{lcccccc}
        \toprule
        & \multicolumn{2}{c}{NRCan LandCover} & \multicolumn{2}{c}{OpenStreetMap} \\
        \cmidrule(lr){2-3} \cmidrule(lr){4-5}
        Method & Trees=15m & Tree=75th pct. & Wood=15m & Wood=75th pct. \\
        \midrule
        High-Rx CAN & 13.5 & \textbf{11.7} & 14.5 & 11.9 \\
        Low-Rx CAN & \textbf{7.8} & 8.3 & 8.8 & 9.8 \\
        \bottomrule
    \end{tabular}
\end{table}

While using other geodata such as NRCan and OSM could potentially enhance environmental classification efforts, as shown in Table \ref{tab:rmse_other}, different classifications may not fully address challenges associated with P.1812. Having four tree categories and assigning different representative clutter heights to each tree type (NRCan) does not necessarily lead to improved results over ESA, which only has one tree category. There are limitations in how these datasets can classify complex environments, and greater complexity does not necessarily guarantee better performance for models such as P.1812.

When discussing the selection of datasets for P.1812 modeling, the chosen dataset should be selected based on the availability and update frequency of geospatial information. ESA offers globally available 2021 data, while with its crowd-sourced nature, OSM, is updated regularly worldwide, making it an excellent resource for capturing recent environmental changes. NRCan, which is updated every five years, is limited to Canada only. Therefore, users should assess which datasets best meet their specific needs, considering the timeliness and coverage. The UK results are not included in Table \ref{tab:rmse_other}, as the NRCan dataset inherently does not provide coverage for that region. For meaningful insights into the UK, it would be critical to identify datasets that are tailored to the region's unique geographical and environmental characteristics.

\section{Conclusion}

This work investigates the impact of geospatial data on the performance of the P.1812 propagation model, emphasizing the importance of geodata in achieving accurate predictions. Tests conducted over million of points found that the ESA dataset, using the default clutter heights, which is consistent with the principles outlined in the ITU-R Recommendation, is well-suited for scenarios across Canada and the UK. Meanwhile, the GCFH dataset performs well in scenarios with higher receivers.

This study also examined the assumption that a higher number of clutter classes or higher resolution of geospatial data would improve model performance. Contrary to expectations, our results show that more detailed inputs or higher granularity do not consistently lead to improved accuracy. Regardless of the land cover dataset used or the quality of its clutter classification, these alone will not automatically improve prediction accuracy. While the geodata can contribute to enhancements under certain conditions, such improvements are not guaranteed across all scenarios.

Further research is needed to refine guidelines for effectively utilizing P.1812, but also for similar ITU-R models, such as ITU-R P.452-18 \cite{P452}, exploring the applicability to higher frequencies. Additionally, future studies could focus on examining the geodata to better understand the factors contributing to variations in their predictions. By assessing various geodata and examining methodologies to create path profiles, this study contributes to improved propagation modeling.

\bibliographystyle{IEEEtran}
\bibliography{references} 

\end{document}